\documentclass[prl,twocolumn,showpacs,amsmath,amssymb,aps]{revtex4}

\usepackage{graphicx}
\usepackage{dcolumn}
\usepackage{amsmath}
\usepackage{bm}

\begin{document}
\title{Electronic Transport in Single-Molecule Magnets  on Metallic Surfaces}

\author{Gwang-Hee Kim$^{1}$}
\email{gkim@sejong.ac.kr}
\author{Tae-Suk Kim$^{2}$}
\email{tskim@phya.snu.ac.kr}
\affiliation{$^1$Department of Physics, Sejong University, Seoul
143-747, Republic of Korea  \\
$^2$School of Physics, Seoul National University, Seoul
151-742, Republic of Korea}
\received{\today}

\begin{abstract}
An electron transport is studied in the system which consists of
scanning tunneling microscopy-single molecule magnet-metal. Due to
quantum tunneling of magnetization in single-molecule magnet,
linear response conductance exhibits stepwise behavior with
increasing longitudinal field and each step is maximized at a
certain value of field sweeping speed. The conductance at each
step oscillates as a function of the additional transverse
magnetic field along the hard axis. Rigorous theory is presented
that combines the exchange model with the Landau-Zener model.
\end{abstract}


\maketitle

 Recently high-spin molecular nanomagnets such as ${\rm Mn_{12}}$ or ${\rm Fe_{8}}$
attracted lots of attention due to observation of quantum
tunneling of magnetization and possible applications in
information storage and quantum
computing\cite{fri,bar,gar,garg,wer99,wer}. These single-molecule
magnets (SMMs) exhibit steps in the hysteresis loops at low
temperature, which is attributed to resonant tunneling between
degenerate quantum states or quantum tunneling of
magnetization(QTM). These unique features of SMMs are the
consequence of long-living metastable spin states due to the large
spin and strong anisotropy
of SMMs.
QTM also made it possible to detect the
interference effect of Berry's phase on the magnetization at each
step while the transverse field along the hard axis is
varied\cite{garg,wer99}.
Novel features of quantum tunneling are
expected to manifest themselves in, if any, other observables.
Especially the effects of QTM on the electronic transport remain
to be explored in both experiments\cite{cor} and theories.

\begin{figure}[b!]
\includegraphics[height=4.5cm]{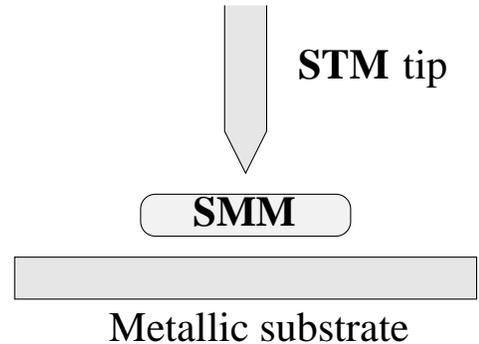}
\caption[0]{Schematic diagram of our model system. A
single-molecule magnet (SMM) is deposited on a metallic surface
and the scanning tunneling microscopy (STM) tip is positioned
right above the SMM. The easy axis of SMMs is directed normal to
the metallic substrate.} \label{fig:geo}
\end{figure}

 In this paper we study theoretically the effects of QTM on
the transport properties of SMMs which are deposited on a metallic
surface with monolayer coverage. Placing the scanning tunneling
microscopy(STM) tip right above one SMM, we compute the electric
current which flows through a SMM when the bias voltage is applied
between the STM tip and the metallic substrate(Fig.
\ref{fig:geo}). We find that the linear response conductance
increases stepwise like the magnetization of a SMM as a
longitudinal magnetic field is increased. The stepwise behavior of
conductance results from the QTM in SMM. The conductance at each
step oscillates periodically as a function of additional
transverse magnetic field along the hard axis. Our theoretical
predictions are not known in the literature as far as we know and
can be tested experimentally.

 When a finite bias voltage is applied between the STM tip and
the metallic substrate, the electrons will tunnel through a vacuum
between the metal surface and the STM tip. Since the STM tip is
placed right above the SMM in our model system, the tunneling
electrons may well be scattered by the large spin of a SMM. Our
model system can be considered as the conventional tunnel junction
with a SMM sandwiched between two normal metallic electrodes. The
metallic substrate and STM tip are conveniently called the
left($p=L$) and right($p=R$) electrodes, respectively. Two
electrodes are described by the featureless conduction bands with
the energy dispersion $\epsilon_{pk}$, ${\cal H}_p =
\sum_{k\alpha} \epsilon_{pk} c_{pk\alpha}^{\dag}
 c_{pk\alpha}^{\phantom{\dag}}$.
The Hamiltonian of the SMM will be introduced later.
Tunneling electrons are modeled by the Hamiltonian\cite{appelbaum,anderson}
\begin{eqnarray}
\label{hamil}
{\cal H}_1
  &=& \sum_{k k' \alpha} \left(
        T_{LR} c^{\dag}_{Lk\alpha} c_{Rk' \alpha}^{\phantom{\dag}}
        + {\rm H.c.} \right) \nonumber \\
  && + \sum_{k \alpha} \sum_{k' \beta} \left(
    J_{LR} c^{\dag}_{Lk\alpha} \vec{\sigma}_{\alpha \beta}
     c_{R k' \beta}^{\phantom{\dag}} \cdot \vec{S}
    +  {\rm H.c.} \right), \label{h1}
\end{eqnarray}
where $\alpha$ and $\beta$ indicate the spin direction of electrons.
The first line represents the direct tunneling between two electrodes,
while the second line describes the tunneling of electrons scattered by
the spin $\vec{S}$ of SMM. Our theory is equally applicable to the
molecular break junction geometry.

 The electric current can be computed using the Keldysh Green's function method
or equivalently the Fermi's golden rule\cite{appelbaum}.
In this paper we study the very weak coupling limit so that
the higher order process like the Kondo effect may be safely neglected.
In this case it is enough to compute the electric current up to
the leading order term.
Using the Fermi's golden rule the electric current can be written as
\begin{eqnarray}
\label{current}
I_{LR}
 &=& e \sum_{m} P_m \sum_{k \alpha} \sum_{k' \beta}
   W_{ L k \alpha m \to R k' \beta m'  }  \nonumber\\
 &&  \times  f(\epsilon_{Lk }^{\phantom{*}} )
   [ 1 - f(\epsilon_{R k' }^{\phantom{*}}) ]
   - ( L k \alpha m \leftrightarrow R k' \beta m').
\end{eqnarray}
Here $W_{i\to j}$ is the transition rate from the state $i$ to
$j$, $f(\epsilon)$ is the Fermi-Dirac distribution function and
$P_m$ is the probability for the SMM to be in the state $S_z = m$.
The leading contribution to the transition rate is given by the
expression
$W_{ i \to j  }
 = \frac{2 \pi}{\hbar} | < j |
{\cal H}_1 | i > |^2 \delta ( E_i - E_j )$, where $i$ and $j$  are
the collective indices denoting the states $\{L k \alpha m \}$ or
$\{R k' \beta m'\}$ and $E_{pk \alpha m} =
\bar{\epsilon}_{pk}+E_m$ with
$\bar{\epsilon}_{pk}=\epsilon_{pk}+\mu_p$ ($p=L,R$). $\mu_p$ is
the chemical potential shift in the electrode $p$ due to the
source-drain bias voltage, and $E_m$ is the energy of the state
$S_z = m$ in the SMM.

 Up to the second order in $T_{LR}$ and $J_{LR}$,
we find the electric current to be
\begin{eqnarray}
I_{LR}
 &=& \frac{2 e^2}{h}  \left[ \gamma_T^{\phantom{\dag}}
    + \langle S_z^2 \rangle \gamma_J^{\phantom{\dag}} \right] V  \nonumber \\
 &&+  \frac{e}{h} \gamma_J^{\phantom{\dag}} \sum_m P_m [S(S+1)-m(m \pm 1)] \nonumber \\
 &&\times  [\zeta(E_m-E_{m\pm 1}+eV )-\zeta(E_m-E_{m\pm 1}-eV) ],
 \label{current2}
\end{eqnarray}
where $\gamma_T^{\phantom{\dag}} (\gamma_J^{\phantom{\dag}})
 = 4\pi^2 N_L N_R |T_{LR}|^2 (|J_{LR}|^2)$ is a measure of the dimensionless
direct (spin-scattered) tunneling rate, $\langle S_z^2 \rangle =
\sum_m m^2 P_m$, $V$ is the source-drain bias voltage given by
$eV=\mu_L -\mu_R$, and $\zeta(\epsilon)=\epsilon/[1-\exp(-\beta
\epsilon)]$ with $\beta^{-1}=k_B T$. The linear response
conductance is then
$G = \frac{2 e^2}{h} \left[ \gamma_T^{\phantom{\dag}}
    + \gamma_J^{\phantom{\dag}} g_s (T) \right]$,
where
$g_s (T) = \langle S_z^2 \rangle + \sum_m P_m [S(S+1)-m(m\pm 1)]
  \eta(E_m-E_{m\pm 1} )
$ with $\eta(\epsilon) = d\zeta(\epsilon)/d\epsilon$.
%
%
We would like to emphasize that only the spin-exchange tunneling
reflects the dynamics of the QTM inside the SMM.

Due to the crystal electric field arising from the structure of a
magnetic molecule, the ground state spin multiplet cannot remain
degenerate.
 The effective Hamiltonian for the ground state spin
multiplet of independent SMMs  such as ${\rm Fe_{8}}$ can be
expanded as\cite{wer99,ras}
\begin{eqnarray}
{\cal H}_{SMM}
 &=&-D S^{2}_{z} + E (S^{2}_{x}-S^{2}_{y}) + C (S^{4}_{+}+S^{4}_{-}) \nonumber \\
 && - g \mu_B^{\phantom{*}} (H_z S_z + H_x S_x), \label{hamil-spin}
\end{eqnarray}
where $S_x$, $S_y$, $S_z$ are three components of the spin
operator, $S_{\pm}=S_x \pm i S_y$, $D$ and $E$ are the
second-order and $C$ the fourth-order anisotropy constants, and
the last term is the Zeeman energy. In the absence of transverse
terms, the energy level of the state $S_z = m$ is $E_{m} = - D m^2
- g \mu_B^{\phantom{*}} H_z m$. When we start with a ground state
$S_z = -S$ corresponding to a large negative longitudinal field,
the level crossing with states $S_z = S-M$ ($M=0,1,2,\cdots$)
occurs at resonant fields, $H_z = H^{(0)}_M = MD/g\mu_B$. When
$H_z = H^{(0)}_{M}$, the two states $S_z = -S$ and $S_z =S-M$ are
degenerate energetically.
Turning on the transverse terms leads to mixing of two degenerate
states, lifts the degeneracy at the resonant fields and results in
the avoided level crossing.

 The scaled conductance $g_s$ can be simplified as
$g_s(M) = S^2+\sum_{n=0}^{M} n P_{S-n}$ at zero temperature by
noting that $E_S <E_{S-1}<...$ and $\eta(\epsilon) =
\theta(\epsilon)$, the step function. In deriving this expression
of $g_s(M)$ it is assumed that the weight transfers from $S_z =
-S$ to $S_z = S,S-1, \cdots$ with increasing longitudinal magnetic
fields.

 To compute the probability, we need to solve the time-dependent
Schr\"{o}dinger equation for the Hamiltonian ${\cal H}_{SMM}$. The
probability is defined as $P_j \equiv \lim_{t\to\infty}
|a_j(t)|^2$ when the wave function is written as $|\Psi(t) \rangle
= \sum^{S}_{j=-S} a_j (t) |j\rangle$. The time-dependent
Schr\"odinger equation for $|\Psi(t)\rangle$ is reduced to the
coupled $2S+1$ differential equations for the coefficient
$a_j(t)$. Recently it was numerically found\cite{ras,rae} that the
two-level approximation can reproduce quite well the results of
the full differential equations. In the ensuing discussion we
adopt the two-level approximation to find an analytic formula of
the probability. The weight transfer is found to occur only
between the states $S_z=-S$ and $S_z=S-M$ at the resonant field
$H^{(0)}_{M}$, for $M=0, 1, 2,..$, until the complete depletion of
the state $S_z=-S$. The amount of such weight transfer depends on
the magnitude of the tunnel splitting or mixing $\Delta_M$ between
two states. At the resonant field $H^{(0)}_{M}$ the full
Hamiltonian ${\cal H}_{SMM}$ is approximated as the effective
two-level model\cite{ras,lan} between the states $S_z=-S$ and
$S_z=S-M$,
\begin{eqnarray}
\cal{H}_{\rm eff}
 &=& \begin{pmatrix}
   - (S-M) g \mu_B c t  & \Delta_M /2 \cr
    \Delta_M /2  & S g \mu_B   c t
   \end{pmatrix},
\end{eqnarray}
where $c$($=dH_z/dt$) is the field sweeping speed.

 Defining $\tau=g\mu_B ct/\Delta_M$ and $\gamma_M=\hbar g \mu_B
c/\Delta^{2}_{M}$, we find the coefficient\cite{ghkim}
around the resonant field $H^{(0)}_{M}$,
\begin{eqnarray}
a_{S-M}^{\phantom{*}} (\tau)
 &=& \sqrt{\lambda_M  \prod^{M-1}_{j=0}F_j  }
  \exp\left[-\frac{1}{4} \left( i\frac{M}{\gamma_M}\tau^2
            + {\pi \lambda_M} \right)
      \right] \nonumber \\
 &&  \times D_{-i\lambda_M-1 } \left[-(1+i) \sqrt{ \alpha_M}  \tau \right],
 \label{asm}
\end{eqnarray}
where $\alpha_M =(2S-M)/(2 \gamma_M)$, $\lambda_M=1 /( 8 \alpha_M
\gamma^{2}_{M})$, $F_j = \exp ( -2\pi \lambda_j )$ and $D$ is the
parabolic cylinder function\cite{gra}. The desired probabilities
are then $P_{S-M}= (1-F_M) ( \prod^{M-1}_{j=0}F_j)$ and
$P_{-S}=\prod^{M}_{j=0}F_j$. Note that $F_j$ and $1-F_j$ denote
the probability for an SMM not to transfer and to transfer from
$S_z=-S$ to $S_z = S-j$ at the $j$-th resonant field,
respectively.

\begin{figure}[t!]
\vskip -0.5cm
\includegraphics[width=9.0cm]{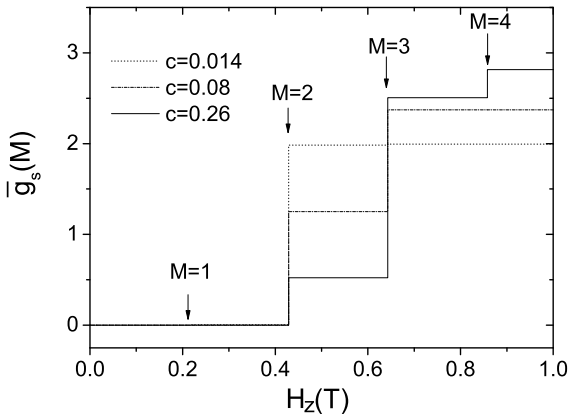}
\vskip -0.5cm \caption[0]{ The scaled conductance $\bar{g}_s(M)$
vs. the longitudinal field $H_z$ at zero temperature for three
typical sweeping speed (T/sec). $M=1,2,3,4$ indicate the positions
of the resonant fields, $H^{(0)}_{M}$= 0.215, 0.429, 0.643, 0.858.
in units of Tesla. } \label{fig:all-res}
\end{figure}

To illustrate the above analytical results with concrete number,
we compute the scaled conductance, $\bar{g}_s$($ \equiv g_s-S^2$)
at zero temperature for an octanuclear iron(II) oxo-hydroxo
cluster of formula ${\rm [Fe_8 O_2 (OH)_{12} (tacn)_6]^{8+}}$
where tacn is a macrocyclic ligand\cite{bar}. We adopt the model
parameters from Refs. \onlinecite{wer99} and \onlinecite{ras}:
$D=0.292$K, $E=0.046$K, $C=-3.2 \times 10^{-5}$K. The tunnel
splitting $\Delta_M$ is calculated for $H_x=0.1 H_z$ at the
resonant field by employing the numerical
diagonalization\cite{ras} or the perturbation method\cite{gar02}.
We obtain qualitatively the same results when $H_x$ has the fixed
value at all resonant fields\cite{twokimtobe}.

 The scaled conductance,
$\bar{g}_s(M) = \sum_{i=1}^M \prod_{j=0}^{i-1} F_j
 - M \prod_{j=0}^{M} F_j$ which is valid for $H_M^{(0)} \leq H_z < H_{M+1}^{(0)}$,
 is displayed in Fig.~\ref{fig:all-res}
for three typical field sweeping speeds. Similar to the
magnetization curve, the scaled conductance is featured with the
stepwise increase as a function of magnetic fields. The jumps in
$\bar{g}_s(M)$ occur at the resonant fields and are caused by the
QTM in SMMs. The step height is very tiny ($\sim 0.318 \times
10^{-4}$) at $H_1^{(0)}=0.215$ T for all three sweeping speeds. At
the second and third resonant fields the step heights are more
pronounced and
 their magnitude depends sensitively on the value of $c$.
Some steps are missing depending on both the sweeping speed and
the resonant fields.

\begin{figure}[t!]
\vskip -0.5cm
\includegraphics[width=9.0cm]{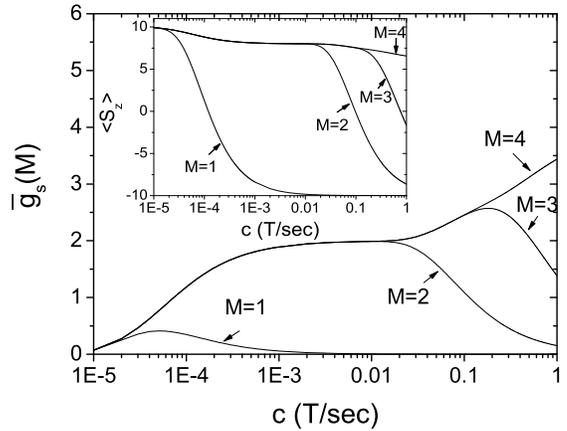}
\vskip -0.5cm \caption[0]{Dependence of $\bar{g}_s(M)$  on the
field sweeping speed $c$ at each resonant field. Inset: the
magnetization $\langle S_z \rangle$ vs. $c$ at each resonant
field. } \label{stepheight}
\end{figure}

 To study in more detail the structure of the steps in the conductance
we plot in Fig.~\ref{stepheight} the scaled conductance
$\bar{g}_s(M)$ at each resonant field as a function of the
sweeping speed $c$. In comparison the magnetization,
 $\langle S_z \rangle = S  - \sum^{M}_{i=1}  \prod^{i-1}_{j=0}F_j
   -(2S -M)  \prod^{M}_{j=0}F_j$, is displayed in the inset.
The magnetization is a monotonically decreasing function of $c$
while the conductance is nonmonotonic and maximized at the
specific value of $c$.
Since the weight transfer, $1-F_j$ at $H_j^{(0)}$,
from $S_z = -S$ to $S_z = S-j$ is monotonically
decreasing with increasing $c$, the magnetization is expected to
decrease with $c$.

Unlike the magnetization, the conductance has contributions only
from the transferred states but not from $S_z = -S$. Since $F_j$
is increasing with $c$, $P_{-S}$ is an increasing function of $c$
while $P_{S-M}$ has the maximum value as a function of $c$. The
conductance $\bar{g}_s(M)$ has the contribution $\delta\bar{g}_s =
M P_{S-M}$ from the $M$-th resonance and is expected to have the
maximum value at some value of $c$. Such a sweeping speed can be
computed approximately as $c^{(\rm max)}_{M} \simeq [ \pi /(2\hbar
g \mu_B )] [\Delta^{2}_{M}/(2S-M)] [\log[M \sum^{M}_{i=0}
\nu_{i}/(1 + \sum^{M-1}_{j=1} \sum^{j}_{i=0} \nu_{i})]]^{-1}$
%
%
where $\nu_{i}=(2S) \Delta^{2}_{i}/[(2S-i) \Delta^{2}_{0}]$.
The values of $c^{(\rm max)}_{M}$ (T/sec) are $5.1 \times 10^{-5}$,
$1.08 \times 10^{-2}$, 0.182 at $M=1, 2, 3$, respectively.
Even though there exists a maximum in the scaled conductance at $M=4$,
the value of $c = 5.16$ (T/sec) lies beyond experimentally meaningful range.
In order to observe the steps in conductance at $M=3$ or $M=4$ resonance,
the sweeping speed should be larger than about 0.01  or 0.1 (T/sec), respectively.

\begin{figure}[t!]
\vskip -0.5cm
\includegraphics[width=9.0cm]{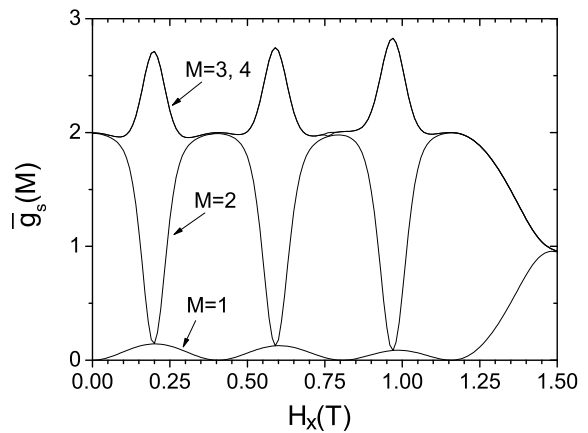}
\vskip -0.5cm \caption[0]{Oscillation of $\bar{g}_M$ as a function of transverse
field for $c=0.014 {\rm T/sec}$.
For this sweeping speed the $M=3,4$ curves are almost identical
except $H_x \simeq 0$, 0.4, and 0.8 T.} \label{osc}
\end{figure}

The conductances at the resonant fields are displayed in Fig.~\ref{osc}
as the transverse field is varied along the hard axis.
Similar to the magnetization the conductance at each resonant field
oscillates with almost the same period of $\sim 0.4$ T.
Such oscillatory conductance faithfully reflects the structure of
the tunnel splittings as a function of the transverse field\cite{twokimtobe}.
The periodic modulation of tunnel splittings by the transverse field
results from the interference between two spin paths of opposite windings
around the hard axis\cite{garg,chu93,vil,wer99}.

 The tunneling splitting is known to vanish at the lattice
of the diabolic fields\cite{vil}. At such fields the tunneling
probability is zero so that the jump in the conductance vanishes.
Depending on the parity of $M$, the oscillations of the
conductance have the different phase. The $M=2$ curve is out of
phase compared to the $M=1,3$ curves. For example, the conductance
for $M=2$ takes on the minimum value at the transverse field where
the conductance for $M=1$ is maximized. This parity behavior
originates from the impossibility of matching an even-valued wave
function with an odd-valued one which gives rise to diabolic
fields. Weak structures around $H_x=0$, 0.4, and 0.8 T for $M=3,4$
curves can be made conspicuous with varying the field sweeping
speed. Though the overall structure of oscillatory conductance
persists, the amplitude of oscillations depends sensitively on the
sweeping speed\cite{twokimtobe}.

 We briefly address the effect of experimentally relevant issues
on our theoretical results. It may be important to consider the
effect of environmental degrees of freedom such as phonons,
nuclear spin and dipolar interaction \cite{gargkim} on the
magnetization process of SMMs. Such interactions make the SMM
relax to the true ground state $S_z=S$ and the relaxation process
helps the magnetization to recover its full stretched value. Since
all the transferred states $S_z=S-M (M=1,2,\cdots)$ lose the
weight to the ground state, we expect that the value of
$\bar{g}_s$ will rise stepwise with increasing field and might
vanish in the end due to the relaxation process. Since the elapsed
time between steps, which is of the order of 10 sec or less for
the typical sweeping speeds (see Fig.~\ref{fig:all-res}), is much
smaller than the relaxation time of magnetization ($\sim 10^{4}$
sec)\cite{bar,gargkim}, we believe that the stepwise behavior of
the conductance can be observed experimentally in the typical
field sweeping speeds.
%
%
The effect of anisotropy in SMMs on the conductance was clarified
in our work. In the absence of anisotropy $g_s = S(S+1)$ so that
the anisotropy in SMMs modifies the conductance by the amount $S$
out of $S(S+1)$. In the case of Fe$_8$ or Mn$_{12}$ $S=10$ so that
the modified conductance is estimated to about 10\% which lies in
the experimentally detectable range.
Possible exchange anisotropy in spin-scattered tunneling can be
addressed\cite{twokimtobe} by considering the ratio,
$a = (J_{LR}^x + J_{LR}^y)^2/4[J_{LR}^z]^2$.
When $a>1$, the conductnce steps are more enhanced than
the isotropic case($a=1$).
For the case of $a < 1$, the steps are reduced or
can be negative depending on the value of $a$.

In summary we studied the current-voltage characteristics of
the STM-SMM-metal system at low temperature.
We found that the quantum tunneling of magnetization (QTM) in SMMs
has a substantial effect on the electronic transport.
The QTM in SMMs leads to the stepwise behavior in the conductance
(just like the magnetization)
when the magnetic field is applied along the easy axis.
Unlike the magnetization the conductance at each resonance
is nonmonotonic with the sweeping speed
and reaches the maximum at some sweeping speed.
In addition, the conductance at the resonant fields
is oscillating as a function of the transverse field applied
along the hard axis.

G.-H.K. was supported by Korea Research Foundation Grant
(KRF-2003-070-C00020). T.-S.K. was supported by Korea Research
Foundation Grant (KRF-2003-C-00038) and grant No. 1999-2-114-005-5
from the KOSEF.


\begin{thebibliography}{10}


\bibitem{fri}
J. R. Friedman \textit{et al.}, Phys. Rev. Lett. {\bf 76}, 3830
(1996); L. Thomas \textit{et al.}, Nature {\bf 383}, 145 (1996).

\bibitem{bar}
A.-L. Barra \textit{et al.}, Europhys. Lett. {\bf 35}, 133 (1996);
C. Sangregorio \textit{et al.}, Phys. Rev. Lett. {\bf 78}, 4645
(1997).

\bibitem{gar}
D. A. Garanin and E. M. Chudnovsky, Phys. Rev. B {\bf 56}, 11102
(1997); V. V. Dobrovitski and A. K. Zvezdin, Europhys. Lett. {\bf
38}, 377 (1997); L. Gunther, {\it ibid.},  {\bf 39}, 1 (1997); E.
M. Chudnovsky and D. A. Garanin, Phys. Rev. Lett. {\bf 87}, 187203
(2001).

\bibitem{wer}
W. Wernsdorfer \textit{et al.},
Nature {\bf 416}, 406 (2002); M. N. Leuenberger and D. Loss, {\it
ibid.}, {\bf 410}, 789 (2001); J. Tejada \textit{et al.},
Nanotechnology {\bf 12}, 181 (2001).

\bibitem{garg}
A. Garg, Europhys. Lett. {\bf 22}, 205 (1993).

\bibitem{wer99}
W. Wernsdorfer and R. Sessoli, Science {\bf 284}, 133 (1999).


\bibitem{cor}
A. Cornia  \textit{et al.}, Angew. Chem. {\bf 42}, 1645 (2003).

\bibitem{appelbaum}
J. A. Appelbaum, Phys. Rev. Lett. {\bf 17}, 91 (1966);
 Phys. Rev.  {\bf 154}, 633 (1967).

\bibitem{anderson}
P. W. Anderson, Phys. Rev. Lett. {\bf 17}, 95 (1966).

\bibitem{ras}
E. Rastelli and A. Tassi, Phys. Rev. B {\bf 64}, 064410 (2001);
{\it ibid.}, {\bf 65}, 092413 (2002).

\bibitem{rae}
H. De Raedt \textit{et al.}, Phys. Rev. B {\bf 56}, 11761 (1997).

\bibitem{lan}
L. Landau, Phys. Z. Sowjetunion {\bf 2}, 46 (1932);
C. Zener, Proc. R. Soc. London, Ser. A, {\bf 137}, 696 (1932).

\bibitem{ghkim} Gwang-Hee Kim (unpublished).

\bibitem{gra}
I. S. Gradshteyn and I. M. Ryzhik, {\it Table of Integrals, Series
and Products} (Academic, New York, 2000).

\bibitem{gar02}
D. A. Garanin and E. M. Chudnovsky, Phys. Rev. B {\bf 65}, 094423 (2002).

\bibitem{twokimtobe} Gwang-Hee Kim and Tae-Suk Kim (unpublished).

\bibitem{chu93}
E. M. Chudnovsky and D. P. DiVincenzo, Phys. Rev. B {\bf 48},
10548 (1993); I. S. Tupitsyn \textit{et al.}, Int. J. Mod. Phys. B
{\bf 11}, 2901 (1997); V. A. Kalatsky \textit{et al.}, Phys. Rev.
Lett. {\bf 80}, 1304 (1998).

\bibitem{vil}
J. Villain and A. Fort, Eur. Phys. J. B {\bf 17}, 69 (2000); A.
Garg, Phys. Rev. B {\bf 64}, 094414 (2001).

\bibitem{gargkim}
A. Garg and G.-H. Kim, Phys. Rev. Lett. {\bf 63}, 2512 (1989); N.
V. Prokof'ev and P. C. E. Stamp, {\it ibid.}, {\bf 80}, 5794
(1998); W. Wernsdorfer \textit{et al.}, {\it ibid.}, {\bf 84},
2965 (2000); L. Bokacheva \textit{et al.}, {\it ibid.}, {\bf 85},
4803 (2000); J. F. Fern\'andez and J. J. Alonso, {\it ibid.}, {\bf
91}, 047202 (2003).

\end{thebibliography}
\end{document}